# Bile dynamics within the biliary tract and microfluidic-based bile component detection: A review

Tao Peng [1], Chenxiao Zhou [2], Zhexin Zhang [3,4], Yingying Liu [3,4], Xiaodong Lin [1], Yongqing Ye [5], Yunlong Zhong [5], Ping Wang [5*], Yanwei Jia [1,3,4,6*]

[1] Zhuhai UM Science & Technology Research Institute, Zhuhai, China

[2] Li Po Chun United World College of Hong Kong, Hong Kong

[3] State Key Laboratory of Analog and Mixed-Signal VLSI, Institute of Microelectronics, University of Macau, Macau, China

[4] Faculty of Science and Technology, University of Macau, Macau, China

[5] Department of Hepatobiliary Surgery, The First Affiliated Hospital of Guangzhou Medical University, Guangzhou, China

[6] MoE Frontiers Science Center for Precision Oncology, University of Macau, Macau, China

* Correspondence: wangping1219@126.com; yanweijia@um.edu.mo

**ABSTRACT:** Bilestones are solid masses found in the gallbladder or biliary tract, which block the normal bile flow and eventually result in severe life-threatening complications. Studies have shown that bilestone formation may be related to bile flow dynamics and the concentration level of bile components. The bile flow dynamics in the biliary tract play a critical role in disclosing the mechanism of bile stasis and transportation. The concentration of bile composition is closely associated with processes such as nucleation and crystallization. Recently, microfluidic-based biosensors have been favored for multiple advantages over traditional bench-top detection assays for their less sample consumption, portability, low cost, and high sensitivity for real-time detection. Here, we reviewed the developments in bile dynamics study and microfluidics-based bile component detection methods. These studies may provide valuable insights into the bilestone formation mechanisms and better treatment, alongside our opinions on the future development of in vitro lithotriptic drug screening of bilestones and bile characterization tests.

**Keywords:** Bile dynamics; Biliary tract; microfluidics; biosensors.

## 1 Introduction

The human biliary system is essential for creating and transporting bile into the duodenum to aid the digestion of fats [1]. It contains the liver, gallbladder, and bile ducts, which are further sorted into cystic, hepatic, and common bile ducts. Acute or chronic biliary diseases such as bilestones, acute acalculous cholecystitis, chronic cholecystitis, and biliary tract malignancies can be life-threatening. Among those biliary diseases, bilestones are the most common, with 10-20% of the global adult population being diagnosed. Although bilestones are not malignant, the continuous recurrence of bilestones has brought immense suffering to a patient and is regarded as a "noncancerous cancer," being a substantial healthcare system burden. Therefore, investigations on how the bilestones are formed and how to effectively prevent and treat bilestones are in high demand [2].

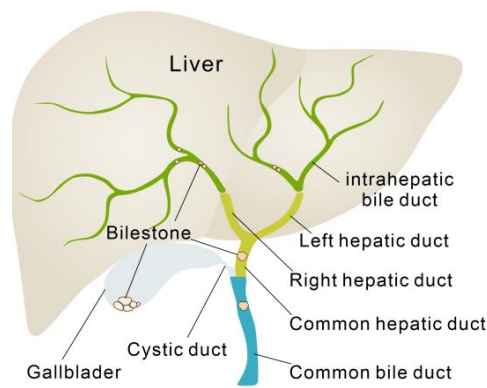

Fig. 1. The schematic view of the biliary tract.

Depending on where the bilestones are formed, they are grouped as gallbladder stones found in the gallbladder, common bile duct stones that occur in the common bile duct, and intrahepatic stones that form in the bile ducts in the liver. Fig. 1 schematically shows the liver structures and various types of bilestones. Depending on the chemical composition, bilestones can be categorized as cholesterol stones, which represent the predominant entity, bilirubin (pigment) stones, and mixed stones [3].

The formation of bilestones in the biliary system is a complex process involving physical and biological factors [4]. Physically, bile duct inflations, bacterial infections, and bile duct narrows would obstruct the bile flow, causing bile stasis and bile component concentration. The concentrated component may trigger bilestone nucleation and growth [5]. Patients with bilestones are often found to have abnormal hydrodynamic data with high biliary resistance. Biologically, genetic abnormalities and liver infections such as hepatitis virus (HPV) and cytomegalovirus (CMV) would cause liver inflation and change its secretions. The blood levels of various bile compositions have been shown to be correlated with the formation of bilestones [6]. Both the physical and biological conditions should be investigated for bilestone formation.

Bile flow dynamics has been the most widely considered factor in exploring the physical conditions of bile ducts. Bile flow dynamics refers to the bile flow velocity and pressure distribution in the biliary tract. However, direct measurement of the bile flow rate in a patient is difficult to monitor in real-time. Nowadays, computational fluid dynamics (CFD) modeling is utilized to provide detailed knowledge about the mechanisms of bile flow dynamics in the biliary tract [7], which provides a means of revealing the relationship between bile dynamics and bilestones.

To investigate the biological factors that can be used to anticipate the risk of bilestones formation and the prognosis after bilestones treatment, blood test of various enzymes is most commonly employed in hospitals. Nevertheless, a central laboratory and a bunch of tubes of blood are needed for different test kits in bulk equipment. To minimize sample consumption, sensitive bile component biosensors have been developed using microfluidics, which handles minute amounts of samples on a chip. The power of microfluidics in integrating sample preparation, reagent mixing, chemical reaction, and results monitoring into one device has brought the potential of lab-on-a-chip for medical diagnostics [8,9].

In this paper, we thoroughly review the literature investigating bile flow dynamics and microfluidic biosensors for bile components. Microfluidics to be used as a platform for organ-on-a-chip is also discussed for in-vitro bilestone studies with the most similar bile duct conditions in vivo. The potential future work that needs to be considered is also discussed at the end of the paper, shedding some light on further investigations of bilestone treatment.

**2 Investigations on bile flow dynamics**

Bile is a greenish-brown, thick substance produced by the liver and stored in the gallbladder. It consists of water, bile acids, cholesterol, phospholipids, bilirubin, and electrolytes. Bile viscosity is an important determinant of local flow characteristics and is essential for studies related to bile dynamics as it affects the flow resistance in the bile ducts [5]. The flow resistance in a duct depends on fluid viscosity, pipe length, and radius, as described in Poiseuille's law [10]:

$$R = \frac{8\eta L}{\pi r^4} \qquad (1)$$

where $\eta$ is the dynamic viscosity, $L$ is the flow path length, and $r$ is the radius of the duct. The resistance in the biliary tract is inversely proportional to the quadratic of the radius, rendering the radius decisive for the flow resistance of the biliary tract and, therefore the bile flow rate. The other factor that linearly affects the flow resistance is the bile viscosity.

### 2.1 Bile rheology

The bile's viscosity characteristics in clinical samples complicated the bile rheology. Bile rheology also depends on sex, pathology type, and shear rate. Most patients' bile samples showed it as a Newtonian fluid whose viscosity is unaffected by shear rate no matter how the flow speeds or shear rates change. However, some patients' bile demonstrated non-Newtonian fluid characteristics that the viscosity varied depending on the shear stress.

Ooi et al. [5] measured bile viscosity with 59 patients who underwent cholecystectomy due to bilestones and found that gallbladder bile viscosity ranged from 1.77 to 8.0 $mPa \cdot s$ at a shear rate of 0.5 to 75 $s^{-1}$. The bile property of 20 subjects was Newtonian fluid, and 22 exhibited shear-thinning properties at a low shear rate and then shear-thickening at a higher shear rate. In 8 subjects, there was shear-thickening, but in 4 other subjects with mucus shear-thining was observed. Through clinical statistics, Reinhart et al. [11] showed that 138 patients (64.5%) had a bile viscosity of 0.7 to 1.1 mPa.s, and 20 patients had a viscosity of >1.4 mPa.s and showed non-Newtonian characteristics, with an exponential increase in viscosity with decreasing shear rate.

It has been found out that bile viscosity differs with biliary locations and between healthy people and patients. Kuchumov et al. have reported that bile in the gallbladder will be more viscous than in the hepatic duct [12]. Coene et al. [13] reported that gallbladder bile from patients with bilestones has a non-Newtonian property of shear thinning at low shear rates, and the dynamic viscosity decreases from 2.5 $mPa \cdot s$ to 1.5

mPa·s when the shear rate increased from 0.1 s$^{-1}$ to 100 s$^{-1}$. Their data also showed that cholecystectomized patients had relatively low bile viscosities without much bile characteristic variation among individual patients. This may be attributed to the resection of the bile bladders in cholecystectomized patients, causing no bile storage and concentrating.

Bile rheology was investigated using both Newtonian and non-Newtonian fluid models. When bile is considered a Newtonian fluid whose viscosity does not vary with shear rate, it is often regarded as water due to the same density and viscosity. Most bile dynamics analysis were on Newtonian fluid [5,14].

When the non-Newtonian model was used in bile dynamic simulation, Carreau's equation[15] and Casson's equation [12] were employed. The constitutive relation for the non-Newtonian fluid flow described by the Carreau's equation is [16]:

$$\eta = \frac{\eta_0 - \eta_\infty}{[1+(a\dot{\gamma})^2]^k} + \eta_\infty \qquad (2)$$

where $\eta_0$ is the viscosity at zero shear rate, $\dot{\gamma}$ is the shear rate, $\eta_\infty$ is the viscosity at infinite shear rate, a is the time constant, and $k$ is the power index.

The Casson's equation gives the relation between stress and shear rate [12]:

$$\begin{aligned} \tau^{1/s} &= \tau_y^{1/s} + \eta(\dot{\gamma})^{1/s} \quad \text{if } \tau \geq \tau_y \\ \dot{\gamma} &= 0 \quad \text{if } \tau \leq \tau_y \end{aligned} \qquad (3)$$

where $\tau$ is the shear stress, $\tau_0$ is the yield stress, and $s$ is the Casson's degree.

**2.2 The geometry property of the biliary system**

The primary function of the biliary tract is to transport the bile. Obtaining the geometry structure of a patient's biliary tract may help predict the bile dynamics guiding the diagnosis of biliary disease and related surgery. For clinical determination of the biliary structure and conditions, medical imaging approaches, including computed tomography (CT), ultrasound (US), and magnetic resonance imaging (MRI), are often utilized [17]. The US method is usually the primary modality for biliary imaging due to its portable, non-invasive, and inexpensive nature. The CT images of the entire biliary tract can be reconstructed at very thin sections for building the biliary system. MRI, unlike CT, can reliably image stones in the biliary tree. The magnetic resonance cholangiopancreatography (MRCP) technique helps diagnose bile duct stenosis and determine the obstruction's location, extent, and severity [17].

With the imaging tools, we have learned the geometry properties of the biliary system. The bile duct is usually a hollow pipe, which consists of four layers, and from inside to outside are the epithelium layer, internal submucosal layer, external submucosal layer, and adventitia [18]. The extrahepatic biliary tract starts with the liver's caudal part and opens into the duodenum [19]. The anatomy of the extrahepatic biliary tract [20] showed that the common bile duct (CBD) length ranged from 1.0 to 7.5 cm, with a mean diameter of 4.0 mm. The cystic duct ranged from 3 to 4 cm long, with a mean diameter of 4.0 mm. The mean length of the left hepatic duct (LHD) was 1.7 cm with a mean diameter of 3.0 mm (±1.08), and the mean length of the right hepatic duct (RHD) was 0.9 cm with a mean diameter of 2.6 mm (±1.2). The cystic duct connects the top of the gallbladder's neck to the common hepatic duct, and anatomy shows that the cystic duct contains concentric folds known as spiral valves of Heister [21]. Deenitchin et al. [22] found that patients with bilestones had longer and thinner cystic ducts than normal individuals, suggesting that flow resistance may be related to cholelithiasis according to Poiseuille's law (Eq. 1).

## 2.3 Investigations of bile dynamics in the biliary tract

Modeling bile dynamics to obtain flow patterns and biliary stresses is vital for revealing the relationship between bile stasis and biliary diseases, such as bilestone and biliary pain. Bile dynamics predictions have been made using simplified and detailed structural models, providing non-invasive mathematical analysis to study bile dynamics facilitating our understanding of bilestone disease. Those models focus on the entire biliary system, the cystic duct, or the hepatic duct.

### 2.3.1 Bile dynamics in the biliary tract

The MRI images can be used to create patient-specific computer-aided design (CAD) biliary tract models [23]. By obtaining a CAD model of the extrahepatic biliary ducts using the three-dimensional (3D) printing technique, in vitro analysis of bile flow and bile duct mechanical properties can be achieved [24]. These studies guide an understanding of the working mechanism and help to analyze bile cholestasis in the biliary tract system.

Kuchumov et al. [12] studied the bile dynamics throughout the biliary system, obtaining velocity and pressure distributions during gallbladder refilling and emptying. In the biliary system, sites of low flow velocity will provide opportunities for subsequent cholestasis. Direct modeling of the biliary system as a whole is often complex. Therefore, Kuchumov et al. [25] decomposed the biliary tract into three parts: extrahepatic biliary tree, gallbladder, and major duodenal papilla. Bile flow in the extrahepatic biliary tree was simulated using the fluid-structure interaction (FSI) algorithm. The Windkessel model was used in their work to simulate the bile flow in the gallbladder, with the bile flow in the major duodenal papilla being assumed to be peristaltic motion. The model can estimate bile flow in the biliary tree. Kuchumov et al. [26,27] further improved their model for pathological bile flow in the major duodenal papilla duct with a calculus. The increasing radius of the calculus may result in a higher pressure drop.

**2.3.2 Bile dynamics in the human cystic duct**

In the biliary tract, the gallbladder collects and discharges bile in the biliary tract through contraction and expansion. The gallbladder empties with a flow rate of about 1 ml/min, as suggested by the ultrasonic graphic imaging data [28], and the flow rate after meal is 2.0-3.0 mL/min [29]. The Reynolds number for bile flow in humans is usually no more than 40, rendering it a steady laminar flow.

The cystic duct is the duct that joins the gallbladder and the common hepatic duct, playing a critical role in a smooth bile flow. Various numerical models have been developed to investigate the flow dynamics in the cystic duct. Ooi et al. [30] reported a one-dimensional numerical model for studying steady flow in the human cystic ducts. In their model, the structure of the cystic duct was simplified to a straight, circular pipe containing staggered baffles. Pressure drop in the cystic duct was predicted with the assumption that higher resistance through the duct would promote bile stasis. The study provides insights into understanding the mechanism of the gallbladder. Li et al. [14] further developed a one-dimensional model to study the dynamics in the biliary system, pointing out that excessive pressure drop may cause incomplete emptying of bile from the gallbladder and cause cholestasis and that the maximum pressure drop ranges from 20 to 100Pa.

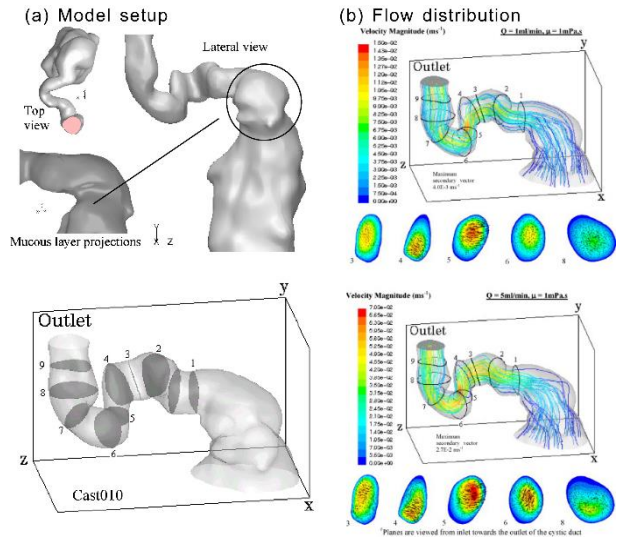

Fig. 2. Bile dynamics study in the cystic duct. Reprinted with permission from [31]. (a) Modeling of the patient-specific cystic duct. (b) The flow distribution in the cystic duct.

It is worth noting that bile shows non-Newtonian characteristics in some clinical results. Non-Newtonian flow usually possesses a higher flow resistance, which enhances cholestasis and triggers difficulties emptying bile from the gallbladder [15]. These 1D models can quickly estimate the pressure drop, but predicting the equivalent stress on the cystic duct is difficult. To address this issue, Al-atabi et al. [32] investigated and visualized the steady bile flow in a two-dimensional (2D) biliary system experimentally, which gives a qualitative indication of higher flow resistance in cystic ducts with bilestones. Then, Al-atabi et al. [31] established a cystic duct model utilizing actual patients' cystic duct lumen with representative anatomical features. They found that the pressure drop was four times that of a linear duct with the same length and diameter due to its convoluting nature. The cystic duct lumen may act as a passive resistor to helps control bile flow out of the gallbladder. The maximal bile velocity and pressure drop depend on cystic duct curvature and radius, as shown in Fig. 2.

For the bile flow study, fluid-structure interaction (FSI) modeling allows for simultaneous bile dynamics and biliary stress distribution prediction. It has also been utilized in the bile flow analysis in the cystic ducts and found that the mechanical stress in the bile ducts is highly correlated with biliary pain [33].

Compared with bile ducts, blood vessels are more widely investigated. The wall shear stress (WSS) in blood vessels is a crucial factor associated with thrombus formation, which is analog to bile stone formation. WSS in the biliary system is generally lower than in the vasculature, similar to thrombogenesis. We expect WSS in the biliary tract would be an excellent factor to predict bilestones formation. However, WSS values in the bile duct are less available in clinical data [16]. More studies are in high demand to validate the relationship between WSS and bilestone formation.

**2.3.3 Bile dynamics in the hepatic duct**

Besides bilestones formed in the gallbladder and cystic duct, most notorious bilestones are found in the hepaonc ducts. Especially, intrahepatic stones formed in the left or right hepatic ducts are notorious for their difficulty in treatment. Kuchumov et al. [16] studied the bile flow pattern in patient-specific bile ducts using FSI simulations to provide information on the bile flow in hepatic ducts. The FSI method provides more accurate results for its thorough prediction of flow and stress distribution. It is found that the wall shear stress (WSS) dominates in the common hepatic and bile ducts, and the maximum WSS increases with increased bile viscosity. When the biliary tract is arranged with stents, the clogging increases the stenosis's shear stress and alters the flow distribution [34]. High flow velocity regions usually correspond to increased WSS magnitude.

When the common bile duct (CBD) is partially obstructed with a bilestone, pressure elevation in the biliary tract will be insignificant. In complete biliary obstruction, the elasticity of the duct wall elevates the duct pressure and limits the deformation of the bile duct [35]. Biliary decompression with a bile duct stent may need to be performed to relieve the obstructed bile flow. Huang et al. [36] conducted FSI simulation analysis, considering the bile, bile duct, and stent models, to obtain the bile dynamics and WSS distribution on the stent and bile ducts. The flow distribution around the stent provides a reference for stent placement in narrow bile ducts and the selection of stent parameters. All these studies provide insights into the fluid mechanics of the biliary system. Tab. 1 lists the details of parameters used in bile dynamics studies.

Tab. 1. Detail of the parameters used for the bile dynamics study.

| Reference | Ducts | Flow | Length | Diameter | Model | Bile property |
|---|---|---|---|---|---|---|
| [15] | Cystic duct | Re:1-40<br>0.12-9.42mL/min | 50mm | 5mm | 1D | Non-Newtonian |
| [14] | Biliary duct | Re: 1-20 | 50mm | 5mm | 1D | Newton fluid |
| [30] | Cystic duct | 0.5-3ml/min<br>Re: 1-40 | 50mm | 5mm | 3D | Newton fluid |
| [31] | Cystic duct | Re:1-40 | 11.2-19.3mm | 2.1-2.6mm | 3D | Newton fluid |
| [32] | Cystic duct | Re: 1-40 | ~ | ~ | 2D | Newton fluid |

## 3 Microfluidic biosensors for detection and test of bile-related component

The formation of bilestones relies on the flow rheology in the presence of stent strictures etc., and bile components, including cholesterol, bile acid, and bilirubin.. Bacterial infection also plays a role in the swelling of the bile duct wall and the bilestone's formation.

The concentration of bile components can be detected using a biosensor, an analytical device that converts biological signals into measurable signals. Microfluidic-based biosensor systems offer multiple advantages over traditional bench-top detection assays with less sample consumption, rapid analysis, portability, and high sensitivity for real-time detection [37]. Existing detection methods for microfluidic biosensors include fluorescence, chemiluminescence, Raman spectroscopy, and colorimetric methods, of which fluorescence and colorimetric methods are commonly used [38]. Microfluidics-based biosensors can detect ultra-low concentrations of bioanalytics with high sensitivity and can be further utilized in the development of patient-friendly and clinically diagnostic-ready devices. Novel microfluidic devices will be powerful tools for studying the biliary system.

Since bile is a "hidden secretion", sampling requires laparotomy or cystic duct cannula, making it difficult to assess the major sources of bile [39]. Little literature can be found on direct bile compositional assays. Most investigations focus on blood testing. In this paper, we review the progress of microfluidics-based platforms for studying bile-related components and bacterial bile duct infection within the past five years.

### 3.1 Microfluidics for cholesterol detection

One of the main functions of cholesterol is to participate in the biosynthesis of bile acids in the liver [39]. Luo. et al. [5] summarized the two conditions for cholesterol bilestone formation, i.e., nucleation of cholesterol crystals and growth of cholesterol crystals. The physicochemical factor of stone formation may be cholesterol supersaturation caused by cholestasis. Increased bile cholesterol concentrations may promote the formation of bilestones through cholesterol crystallization. Although commercial kits for cholesterol determination are readily available, the number of analytical steps and reagents involved in the process makes it tedious and time-consuming [40].

The commonly used microfluidic electrochemical biosensors allow simple, low-cost, and rapid detection of cholesterol in solution. The principle of electrochemical biosensors is that the chemical reaction between the immobilized biomolecule and target analyte produces or consumes ions or electrons, affecting the solution's measurable electrical properties, such as electric current or potential [41]. Electrochemical detection of cholesterol is usually based on enzymatic reactions. Therefore, cholesterol oxidase must be coated on the working electrode to achieve specific cholesterol binding. High sensitivity detection and real-time monitoring of cholesterol can be achieved by measuring the electrical signal changes in the solution. The reported materials for working electrodes are carbon nanotubes [42] and nickel oxide films [43]. Watanabe et al. [44] proposed non-enzymatic determination of cholesterol under alkaline conditions using an electrochemical microfluidic device using the redox pair Ni(II)/Ni(III) as a mediator. The sensor avoids interference from other electroactive biomolecules in the sample.

For cost-effectiveness, three-dimensional microfluidic and paper-based analytical devices (3D-μpad) have recently gained much attention. 3D-μpad was developed using the colorimetric method [44,45] and chemiluminescence [46] to detect cholesterol. Baek et al. [45] report a 3D-μPADs platform with a smartphone for detecting multiple biomarkers, including cholesterol. The colorimetric signal was generated with the enzymatic oxidation reaction of coloring reagents, peroxidase, and $H_2O_2$. Li. et al. [46] developed a 3D μPAD with high-precision temporarily resolved chemiluminescence (CL) emissions to determine cholesterol, as shown in Fig. 3a.

Román-Pizarro et al. [47] immobilized enzymes on magnetic nanoparticles, and the magnetically retained enzyme microfluidic microreactor enabled micro-scale hydrolysis and oxidation of cholesterol, and the monitoring of fluorescence decreasing enabled quantification of cholesterol concentration. Using magnetic nanoparticles to immobilize the enzyme can reduce reagent consumption and enhance the analytical signal. These studies indicate that microfluidic biosensors can be a point-of-care platform for realizing low-cost, rapid, simple, and highly sensitive cholesterol detection in biosamples.

**3.2 Microfluidics for bile acid detection**

Bile acids are products of cholesterol degradation and metabolism, working as biological detergents that promote intestinal digestion of lipids and fat-soluble vitamins [48]. Bile acids are commonly associated with liver function. Healthy individuals and patients with liver diseases clearly differ in bile acid concentrations. Bile acids are biomarkers to diagnose bilestones, hyperlipidemia, and other liver-related diseases [49,50].

In recent years, liquid crystal (LC) droplets combined with microfluidics have been developed to detect bile acids, but these methods usually require complex surface modification processes [51,52]. Coupled biochemiluminescence (BL–CL) enzymatic reactions [53] can also be used for bile acid biosensing with increased sensitivity and limited of detection (LOD). Cedillo-Alcantar et al. [54] proposed an automated microfluidics platform for detecting bile acid in microliter droplets. Automation of droplet generation frequency and size by computer-controlled pneumatic valves improves the detection limit of bile acids in droplets. Han. et al. [55] demonstrated a monodisperse 4-cyano-4′-pentylbiphenyl (5CB) LC droplet-based microfluidic chip, generating and capturing monodisperse LC droplets. The concentration of bile acids was quantitatively correlated by monitoring the collapse of 5CB droplets. These miniaturized instruments have significant potential for the quantitative and rapid detection of bile acids.

Table. 2. The detection of bile components in microfluidic biosensors.

| Analyte | Reference | Method | Range | LOD | Time |
|---|---|---|---|---|---|
| Cholesterol | Kaur et al. [43] | Electrochemical | 0.12-10.23 mM | 0.1 mM | ~ |
| | Watanabe et al. [44] | Electrochemical | 0.1-50 μmol/L | 0.083 μmol/L | ~ |

| | Baek et al. [45] | Colorimetric | 0-10 mmol/L | 0.3 mmol/L | <10 min |
| --- | --- | --- | --- | --- | --- |
| | Li et al. [46] | Chemiluminescence | 2.9-6.0 mM | 6 μM | ~ |
| | Román-Pizarro et al. [47] | Optical | 0.005-10 mMol/L | 1.1 μmol/L | ~ |
| Bile acid | Cedillo-Alcantar et al. [54] | Fluorescence | 0-150 μM | 2.1 μM | ~ |
| | Han et al. [55] | Optical | 10-50 μM | 1 μM | < 4 min |
| Bilirubin | Tan et al. [56] | Colorimetric | 0-30 mg/dL | 1.2 mL/dL | < 10 min |
| | Bandara et al. [57] | Colorimetric | 1-1.5 mg/dL | ~ | ~ |
| E.coli | Yao et al. [58] | Chemiluminescence | $2\times10^2$ - $10^8$ CFU/mL | 130 CFU/mL | 1.5h |
| | Sun et al. [59] | Impedance | $10^1$ -$10^5$ CFU/mL | 12 CFU/mL | 2 h |
| | Zheng et al. [60] | Colorimetric | 50 - $5\times10^8$ CFU/mL | 50 CFU/mL | 1 h |
| | Shang et al. [61] | Fluorescence | $10^2$ -$10^8$ CFU/mL | 10 CFU/mL | 2.5 h |

**3.3 Microfluidics for bilirubin detection**

Bilirubin is an essential component of bile, and bilirubin is usually found in the biliary tract as unconjugated bilirubin (UCB) or conjugated bilirubin (CB) [62]. Conjugated bilirubin in the liver is water-soluble, excreted to the biliary tract and cystic duct, and then passed to the duodenum. CB may hydrolyze to unconjugated bilirubin in the presence of bacteria in the biliary tract [3], which can be dissolved in fat. The average concentration of bilirubin in the blood of a healthy person lies in the range of 0.3–1.9 mg/dL with conjugated bilirubin (0.1– 0.4 mg/dL), unconjugated bilirubin (0.2– 0.7 mg/dL) [63]. Conventional bilirubin assays include the Diazo and vanadate oxidase methods [64].

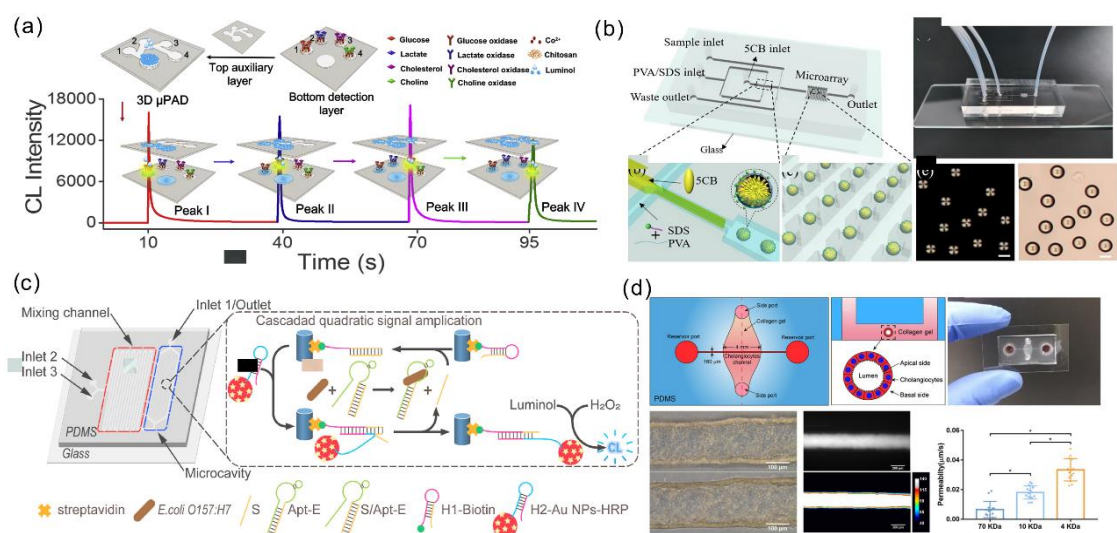

Fig. 3. Microfluidic devices for bile component detection. (a) Schematically illustration for fabrication of 3D μPAD for multiplexed CL analysis. No.3 is the cholesterol

detection zone. Reprinted with permission from [46]. (b) Schematic illustrations of droplet-based microfluidic for bile acid determination. Reprinted with permission from [55]. (c) Microfluidic biosensor for E. coli detection. Reprinted with permission from [59]. (d) The bile duct organ on a chip. Reprinted with permission from [65].

Tan et al. [56] developed a tape-paper sensing method for detecting the total bilirubin levels based on colorimetric diazotization. The tape-paper sensor is a low-cost, fast, and user-friendly device for measuring the total bilirubin levels. Thompson et al. [66] developed a simple, reagent-less quantification of total bilirubin based on microfluidic photo-treatment using a polyester chip, providing rapid results. There are also some microfluidics-based assays for UCB and CB. Bell et al. [67] developed a paper-based potentiometric sensing method for determining free bilirubin in serum, which exhibited a response range of 5.0–0.10 mM and a requirement of 15 μL sample. Bandara et al. [57] reported a membrane-based wicking microfluidic device integrating colorimetric assay chemistries to achieve semi-quantitative detection of CB in the blood sample.

**3.4 Microfluidics for *E. coli* bacteria detection**

Bilestones are associated with bacteria in the bile. For example, *E. coli* produces β-glucuronidase, phospholipase A1, and conjugated bile acid hydrolase in the biliary system that promotes the formation of free bilirubin, which eventually combines with calcium salts to form brown pigment stones [3]. Detecting *E. coli* from bile is important for taking precautions against bilestones.

Sun et al. [59] developed a microfluidic chemiluminescence biosensor based on a multiple signal amplification strategy, and the chip contains an upstream mixing unit for mixing luminol and $H_2O_2$, and downstream micropillars for capturing bacteria and catalytic hairpin assembly to amplify the signal. The method achieves high sensitivity and better selectivity to target bacteria detection. Yao et al. [58] proposed a microfluidic impedance biosensor combined with the immune magnetic nanoparticles (MNPs) for target separation of *E.coli*, gold nanoparticles modified by the aptamers against *E. coli* and the urease were incubated with the magnetic bacteria by measuring the impedance of the catalyst, the microfluidic device allows rapid, sensitive bacteria detection, as shown in Fig. 3c. Zheng et al. [60] proposed a method to correlate *E.coli* concentrations with color changes by smartphone detection of Au nanoparticles (AuNPs) indicators. Magnetic nanoparticles modified with the capture antibodies against *E. coli*, polystyrene microspheres (PSs) modified with the detection antibodies against *E. coli,* and catalases were mixed with the target *E. coli* sample in the microfluidic chip to form MNP-bacteria-PS complexes. The complexes were captured and blended with AuNPs using the external magnetic field. The aggregation of AuNPs was triggered through the crosslinking agents, resulting in the color change, and was detected by a smartphone to determine the amount of bacteria. Researchers particularly favor colorimetric detection methods because they allow the results to be observed directly with the naked eye. The critical challenge is to translate biometric signals into color changes.

Shang. Et al. [61] developed a portable microfluidic sensor utilizing a finger-driven chip containing three units: immunomagnetic separation, nucleic acid extraction and purification, and signal detection. Samples are captured and immobilized in the flow channel employing a mechanical trap, allowing further in situ observation and detection. Antibody-modified MNPs were used to sort bacteria, bacterial DNA was captured by silica-coated MNPs, and quantification of bacteria was achieved based on fluorescence detection. All these merits of this microfluidic biosensor, such as rapid, sensitive, and reliable analysis, made it very promising for detecting *E. coli* and using it for bile tests. Tab. 2 lists the details of parameters used in microfluidic biosensors.

**4 Microfluidics bile duct organ on a chip for bile flow study and test**

Traditional 2D monolayer cell culture systems are the simplest and most cost-effective method for culturing human and animal adherent cells. However, there are many differences between the 2D model and the in vivo environment, such as mechanical, physiology, and hydrodynamic aspects [68]. 3D culture systems address these limitations by controlling various parameters to design microfluidic devices that generalize the characteristics of multiple cellular microenvironments [69]. Advantages of microfluidic cell culture systems include dynamic flow conditions and mechanical stimuli within microchannels that mimic *in vivo* conditions.

Organ-on-a-chip systems are microfluidic devices for culturing living cells in micrometer chambers with continuous flow conditions to generalize the physiological functions of tissues and organs [70]. These systems can provide various *in vivo*-like mechanical deformations, including fluid shear stress, cyclic strain, and compression at physiological levels.

Advanced microfluidics and high-resolution microscopy make it possible to accurately simulate physiological systems and gain insight into the behavior of complex biliary tract systems. Compared to conventional macroscopic systems, microscale systems can better simulate the biliary environment and study the dynamics of disease and drug activity in detail by reproducing organ-tissue interactions. Nakao et al. [71] presented a microfluidic cell culture device in which aligned hepatocytes cultured by microfluidics gradually self-organize and form bile ducts along the hepatic cord-like structure, which can be used for drug screening and toxicological studies in the biliary system. Du et al. [65] proposed a bile duct organ on a chip. The bile duct microenvironment is formed by growing cholangiocytes within microfluidic channels to form a dense bile duct epidermal construct. The organ-on-a-chip can functionally mimic the natural bile duct environment and provides a valuable tool for bile testing, as shown in Fig. 3d. The microfluidic system provides high-quality control and precise generalization of different aspects of the bile duct model.

## 5 Conclusions

This paper reviews the bile dynamics in the biliary tract and the recent development of microfluidic biosensors for detecting bile components. There is relatively little literature on bile dynamics studies compared to blood flow. However, a study on bile flow and bile component is also essential, as bile dynamics data and bile composition levels help reveal biliary tract diseases. Previous studies focused more on biliary structures and less on biliary strictures. Nevertheless, it has been clinically found that patients containing bilestones often have biliary strictures. The bile dynamics after strictures still deserve to be investigated.

Advanced 3D imaging and reconstruction techniques provide a convenient way to construct geometric models of the biliary tract. Real-time patient-specific analysis of the bile dynamics will provide essential data for clinical reference. Meanwhile, thanks to the development of microfabrication technology, it is possible to fabricate complex 3D and tiny 2D structures. The future application of microfluidic platforms for rapid and precise sensing of bile components will bring new tools for clinical applications and immediate diagnosis. There is still a need to develop novel microfluidic-based biochip and organ-on-a-chip platforms for in vitro bilestone-dissolving drug screening tests or biliary function tests, which can be translated into the commercial use of point-of-care tests (POCT). Combining hydrodynamics, biology, histology, and microfluidics for biliary system analysis will further empower and uncover more clinical implications.

**Acknowledgments**

This research is financially supported by the China Postdoctoral Science Foundation (Grant No. 2023M734087), the Macao Science and Technology Development Fund (FDCT) [FDCT 0029/2021/A1, SKL-AMSV (UM)/2023-2025]; University of Macau [MYRG2020-00078-IME, MYRG-GRG2023-00034-IME]; Zhuhai Huafa Group [HF-006-2021]; and Guangzhou Science and Technology Plan Project (Grant No. 202102010251).

**Declaration of competing interest**

The authors declare no conflict of interest.